# Graphene-based chemiresistive gas sensors

M. Spasenović, S. Andrić, T. Tomašević-Ilić

*Abstract* - Gas sensors are an indispensable ingredient of the modern society, finding their use across a range of industries that include manufacturing, environmental protection and control, automotive and others. Novel applications have been arising, requiring new materials. Here we outline the principles of gas sensing with a focus on chemiresistive-type devices. We follow up with a summary of the advantages and use of graphene as a gas sensing material, discussing the properties of different graphene production methods. Finally, we showcase some recent results that point to novel applications of graphene-based gas sensors, including respiration monitoring and finger proximity detection.

## I. INTRODUCTION

Gas sensors are devices that are ubiquitous in many industries, including manufacturing, automotive, building safety and others. Emerging applications such as widely available air quality sensors and wearable devices are exerting a pull on the development of devices from novel materials, that are often thin and flexible. Nanomaterials play an important role in fabricating thin, flexible sensors, due to their ease of manufacturing, scaling and application to a substrate, versatility, and favorable electronic properties such as efficient charge transport.

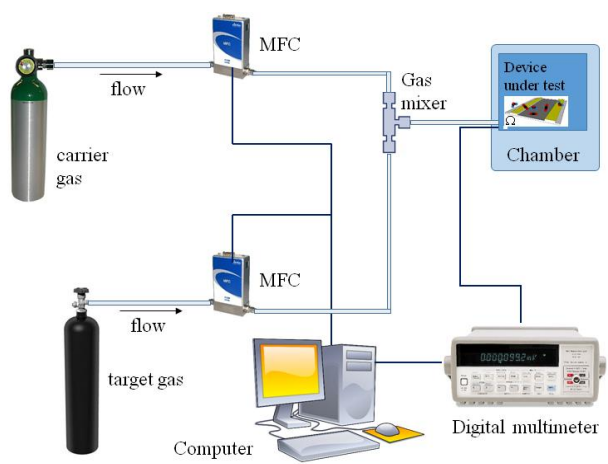

Fig. 1. Experimental setup for measuring response of gas sensors.

M. Spasenović and Stevan Andrić are with the Center for Microelectronic Technologies, Institute of Chemistry, Technology and Metallurgy, University of Belgrade, Njegoševa 12, 11000 Beograd, Serbia, e-mail: spasenovic@nanosys.ihtm.bg.ac.rs.
T. Tomašević-Ilić is with the Institute of Physics Belgrade, University of Belgrade, Pregrevica 118, 11080 Beograd, Serbia.

To characterize and calibrate the operation of a gas sensor, the sensor must be placed in a controlled environment where key parameters are regulated. An example of an experimental setup for testing gas sensors is depicted in Fig. 1. The gas sensor (device under test, DUT) is placed in an environmental chamber. Often, such chambers are custom made by the sensor developer to satisfy requirements of the particular geometry and measurement parameter space. For example, if the sensor is cross-sensitive to temperature or humidity, these parameters must be controlled or at least monitored in the chamber [1].

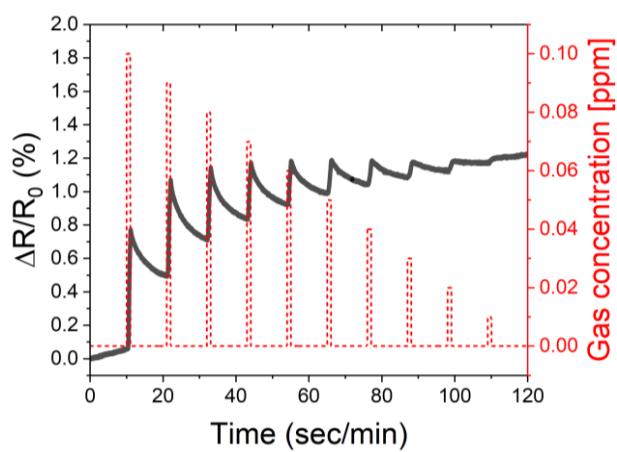

Fig. 2. Typical response of gas sensor to injection of known concentration of analyte. The red dashed line indicates injected gas concentration. The solid black line indicates resistance measured across two sensor terminals.

The DUT is contacted with wires that feed through the walls of the chamber and are connected to a measurement instrument, such as a digital programmable multimeter (DMM). The electrical characteristics of the DUT are measured as the environment is varied in a controlled manner. A constant flow of gas is introduced to the chamber via a gas inlet that is connected to mass flow controllers (MFC) or other gas flow control devices. For sensors boasting high sensitivity and accuracy of detection in the ppm or ppb range, the flow will consist mostly of an inert carrier gas, and a small percentage of the target analyte gas. The two gases are most often mixed in a simple T-shaped mixer prior to entering the detection chamber. MFC and DMM operation can be controlled and synchronized from a computer.

Fig. 2. depicts a common measurement protocol and sensor response of a chemiresistive gas sensor. The change in resistance, compared to measured resistance of the unexposed sensor, is monitored in time as the analyte gas flow is tuned to achieve different target concentrations. In the example shown, a short burst of analyte is introduced every 11s at varying concentrations, starting from the highest. The sensor responds with an increase in resistance at burst onset, followed by a slow decrease when analyte flow is switched off and only carrier gas is flown over the sample. In the case of 2D material-based sensors, there is often a residue background drift of the measured response, which can be filtered out by signal differentiation [2].

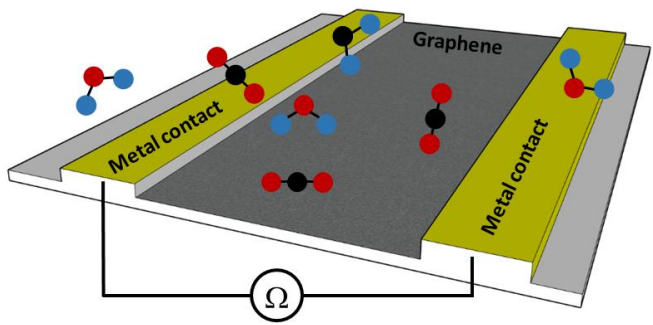

Fig. 3. Graphene-based chemiresistive gas sensing device. Graphene is laid on a substrate with metal contacts and exposed to a gas atmosphere.

Graphene is increasingly being investigated as an active material for gas sensing. Several physicochemical principles are commonly utilized to make gas sensors, including calorimetric sensors, optical sensors, electrochemical sensors, and chemiresistive sensors. Of these types, chemiresistive are the most versatile, and are easily integrated in a planar configuration. Chemiresistive sensors consist of a minimum of two electrical terminals connected to the active sensing layer. Fig. 3. depicts a graphene-based chemiresistive gas sensor. The graphene sits atop a substrate and is contacted with two planar metal contacts. Gas molecules in the vicinity of graphene can interact with the sensor, which is reflected in a changing electrical resistance between the two contacts with changing gas concentration or type.

In the following sections, we discuss the properties of several graphene fabrication techniques as they relate to use in gas sensing. We sketch the role that defects play in the interaction of graphene with gases. Finally, we demonstrate the application of chemiresistive graphene gas sensors to novel sensing technologies.

## II. Graphene-based gas sensors

The chief properties required from a gas sensing material are efficient charge transport to a substrate or high carrier mobility, a large surface area, and an abundancy of active reaction sites. Various forms of graphene fulfill these requirements to different extents. In Fig. 4. we show the most common forms of graphene available in research laboratories and on the market. Micromechanical exfoliation is the oldest method of making graphene, which was utilized by Geim and Novoselov in their seminal paper [3] that sparked the graphene revolution. The method is based on mechanical cleavage of graphite flakes and separation of layers with adhesive tape. Thinned down graphite layers remain on the tape. The process is repeated several times, until visual inspection indicates the presence of very thin graphite layers on the tape. The tape is subsequently applied to a rigid substrate, such as Si/SiO2. After tape removal from the substrate, careful optical microscopy may reveal flakes of few-layer graphene, including single layers. One such flake is depicted in the optical micrograph of Fig. 4a. Typical lateral flake sizes are on the order of several tens of micrometers, whereas flakes are few and far apart on the substrate. This method yields graphene of the highest quality, with very few defects in the basal plane.

The main principle of interaction of micromechanically exfoliated graphene with gas species is through gas adsorption and desorption processes on the basal plane, i.e. physisorption. The adsorbed molecules change the local carrier concentration in graphene, which exerts changes in resistance. With extreme fabrication and noise control, mechanically exfoliated graphene has been used to achieve single-molecule sensitivity [4].

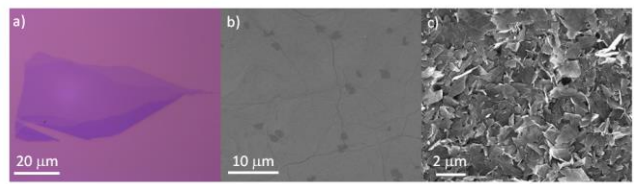

Fig. 4. Different types of graphene. a) depicts an optical micrograph of micromechanically exfoliated graphene; b) depicts a scanning electron micrograph (SEM) of CVD graphene (image courtesy of Graphenea); c) depicts a SEM of liquid phase exfoliated graphene.

The structure of micromechanically exfoliated graphene is similar to that of graphene grown by chemical vapor deposition (CVD). A scanning electron micrograph of CVD graphene is depicted in Fig. 4b. CVD graphene is grown on a catalyst substrate, such as a smooth copper foil, by placing the substrate in a chamber where temperature and growth precursors are controlled. Growth starts from seeds on the substrate and continues until the process is stopped or a continuous film is formed across the substrate. Growth from multiple seeds results in numerous grain boundaries. Because the growth takes place at highly elevated temperatures (>1,000°C), the substrate and the graphene undergo a shrinking and expanding process, respectively, during cooling of the chamber. Due to a mismatch of thermal expansion coefficients of graphene and the substrate, wrinkles inevitably form on the graphene surface [5]. Hence, CVD growth results in graphene that is continuous over a large area, with some structural defects such as wrinkles and grain boundaries. Compared to micromechanically exfoliated graphene, CVD graphene may contain additional binding sites at these defects, which provide a channel for chemisorption in addition to physisorption that takes place on the surface of the basal plane. It was found that polymer residue on graphene that remains after lithographic processes enhances the response of graphene gas sensors, with a surprisingly weak intrinsic response [6].

Graphene that is obtained from reduction of graphene oxide or from liquid phase exfoliation has the morphology of nanoplatelets – thin flakes of graphite or graphene. These flakes can have thicknesses ranging from a single layer of graphene up to tens of layers, and lateral sizes between several tens of nanometers and tens of micrometers. Such graphene is usually obtained in solution, which contains a range of nanoplatelet sizes. There exist several different methods for drawing nanoplatelets from solution into a film, such as spray coating, drop casting, spin coating, inkjet printing, and Langmuir-Blodgett assembly [7]. A film assembled from such nanoplatelets may look like that depicted in Fig. 4c.

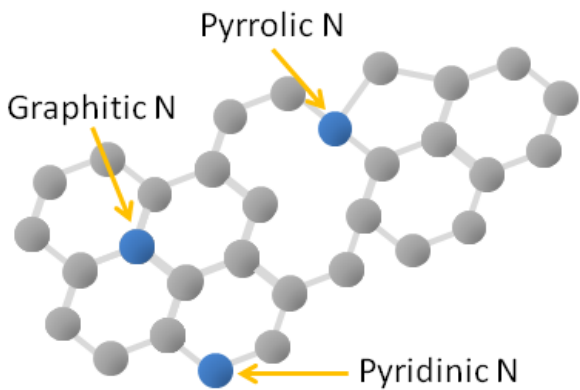

Fig. 5. A sketch of graphene structure and some commonly found defects.

Although films that are made from nanoplatelets typically show poor electrical conductivity compared to films made of CVD graphene, their response to gas exposure is different and may be stronger than the response of CVD graphene. A strong change of electrical properties upon exposure to an analyte in these materials can be attributed to the presence of reactive defect sites. As an example, nitrogen atoms can bind to graphene at various positions in the crystal lattice, as shown in Fig. 5. When a nitrogen atom simply substitutes a carbon atom in the basal plane, the substitution is termed graphitic nitrogen. When the nitrogen replaces carbon at a crystal defect site, one obtains pyrrolic nitrogen. Finally, when nitrogen replaces carbon at the edge of graphene, it is termed pyridinic nitrogen. Each of these species results in different electronic behavior, also affecting chemical reactivity of the graphene. It was observed that certain species such as water vapor will predominantly bind to edges, with a more pronounced effect in thicker platelets, whereas dry oxygen preferably reacts with the basal plane. Even in the case of micromechanically exfoliated graphene that is relatively free of edges, the reaction to water vapor is much stronger than to dry oxygen, which indicates high edge reactivity [8]. Similarly, it was observed that edges are the dominant defect type in liquid-phase exfoliated films and that oxygen predominantly binds to these edges upon exposure to an oxygen plasma [9]. Researchers have also created edges by damaging graphene on purpose, to increase its reactivity to gases [10].

As a result of an abundancy of edges, films that are formed as a continuous sheet of interconnected nanoplatelets tend to react more strongly to the presence of gases than films that are constituted of continuous sheets with very few edge defects. At least one work has shown that films made from liquid-phase exfoliated graphene experience a significant drop in sheet resistance upon short exposure to an oxygen environment [9], as in Fig. 6. The value after 5 minutes of exposure can be less than 50% of the starting value, which is not the case for CVD graphene. The hypothesis presented in the referenced work is that oxygen binding to the edges of nanoplatelets dopes the graphene, causing a decrease in measured resistance. In contrast, CVD graphene reacts to treatment by a slight increase in sheet resistance, which is due to graphene lattice damage caused by oxygen ejecting carbon atoms. After 10 minutes of treatment, the damage becomes visually perceptible, whereas in the case of LPE graphene no damage will occur until all edges are saturated.

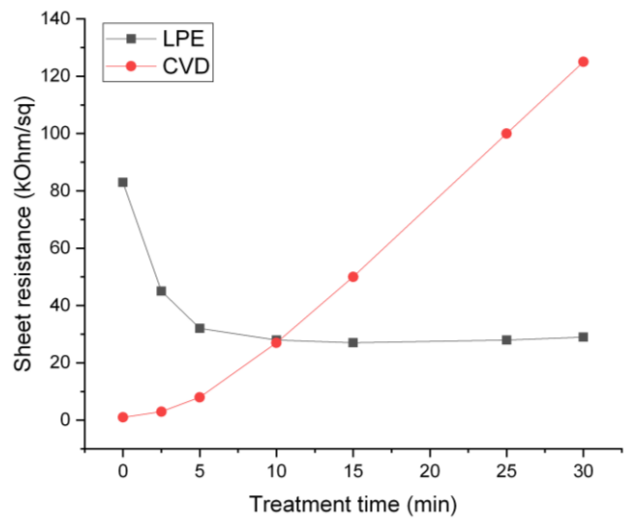

Fig. 6. Typical response of two types of graphene to a chemical treatment. The line with the red circles indicates response of CVD graphene. The line with the black squares indicates response of LPE graphene.

The strong sensitivity of LPE graphene to environmental gases can be exploited to produce gas sensors. In particular, such graphene has been used to detect $NO_2$ [11], [12] with ppm sensitivity, while theory predicts that graphene containing an abundancy of defect sites would react strongly to CO, NO, and $NO_2$[13].

Recently, it was found that LPE graphene is highly sensitive to changes in relative humidity of the ambient [14]. Changes in relative humidity can exert a change in measured resistance of chemiresistive graphene gas sensors by up to 10%. It was also found that the response to humidity is much stronger than the response to other constituents of air, which makes this sensor ideal for open-air use as a humidity detector. Surprisingly, and in contrast to what has been observed in the case of $NO_2$ detection with LPE graphene [12], the response to humidity changes is extremely fast, on the order of 30 ms. Such extremely fast response opens avenues towards the use of these sensors in situations in which humidity changes more rapidly than can be detected with conventional humidity sensors, that are often based on the principle of humidity-induced swelling of a material and measuring a corresponding capacitance change. Indeed, it has been demonstrated that the ultrafast LPE graphene-based sensor can be used to monitor human respiration in real time (Fig. 7). The sensor was made by depositing LPE graphene, using Langmuir-Blodgett self-assembly, onto a pair of interdigitated metal contacts premade on a ceramic substrate. Breathing on the sensor resulted in real-time variation of the measured two-terminal resistance. Another novel application demonstrated in the same publication is real-time sensing of finger proximity detection. The application relies on the principle of detecting the cloud of moisture that exists in the vicinity of human skin. Real-time detection of finger proximity could be used to make novel devices, such as touchless control panels.

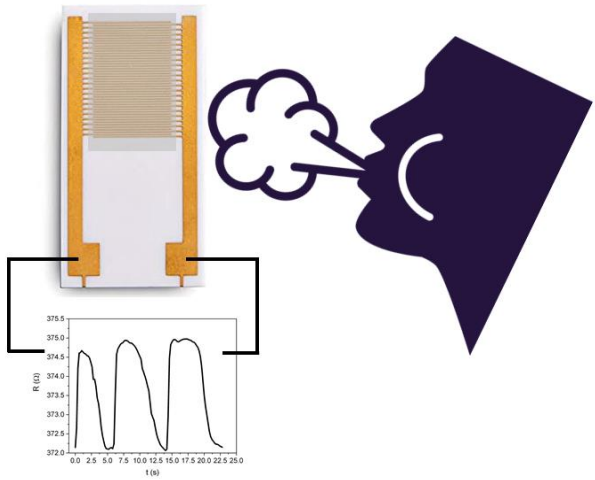

Fig. 7. The use of a chemiresistive graphene-based sensor as a respiration monitor. Resistance between two terminals changes in response to a person breathing on the sensor.

## III CONCLUSION

To conclude, liquid-phase exfoliated graphene presents an excellent opportunity to create practical gas sensors. Although theory predicts strong sensitivity to CO, NO, and $NO_2$, experiments have also shown a very strong response to changes in relative humidity, pointing to the potential for detection of other gases. Furthermore, the utlrafast response to humidity can be utilized to create novel devices, which may prove useful in particular for wearable biometric sensing.

## ACKNOWLEDGEMENT

We acknowledge the Science Fund of the Republic of Serbia through the PROMIS funding scheme, project Gramulsen #6057070. This work was financially supported by the Ministry of Education, Science and Technological Development of the Republic of Serbia (Grant No. 451-03-68/2020-14/200026) and by the Institute of Physics Belgrade, through grants by the Ministry of Education, Science, and Technological Development of the Republic of Serbia.